\PassOptionsToPackage{table,dvipsnames}{xcolor}
\documentclass[sigconf]{acmart}

\usepackage{amsfonts}          
\usepackage{subcaption}
\usepackage{siunitx}
\usepackage{makecell} 
\usepackage{multirow}
\usepackage{soul}
\usepackage{enumitem}

\usepackage[ruled,vlined,linesnumbered]{algorithm2e}

\newcommand{\Eqref}[1]{Equation~\ref{#1}}

\definecolor{yellow}{HTML}{f4c342}
\definecolor{GreenDark}{HTML}{5e813f}
\definecolor{GreenMid}{HTML}{7eab56}
\definecolor{GreenLight}{HTML}{cbdfb9}

\AtBeginDocument{%
  }

\copyrightyear{2025}
\acmYear{2025}
\setcopyright{acmlicensed}\acmConference[ICAIF '25]{6th ACM International Conference on AI in Finance}{November 15--18, 2025}{Singapore, Singapore}
\acmBooktitle{6th ACM International Conference on AI in Finance (ICAIF '25), November 15--18, 2025, Singapore, Singapore}
\acmDOI{10.1145/3768292.3770408}
\acmISBN{979-8-4007-2220-2/2025/11}

\author{Junyi Mo}
\email{junyi.mo@stern.nyu.edu}
\orcid{0009-0001-2285-7019}
\affiliation{%
  \institution{New York University}
  \city{New York}
  \state{New York}
  \country{USA}
}

\author{Jiayu Li}
\email{jl15681@stern.nyu.edu}
\orcid{0009-0002-6625-6167}
\affiliation{%
  \institution{New York University}
  \city{New York}
  \state{New York}
  \country{USA}
}

\author{Duo Zhang}
\email{dz2349@nyu.edu}
\orcid{0009-0005-5951-0851}
\affiliation{%
  \institution{New York University}
  \city{New York}
  \state{New York}
  \country{USA}
}

\author{Elynn Chen}
\email{elynn.chen@stern.nyu.edu}
\orcid{0000-0002-7599-1828}
\authornote{Corresponding author.}
\affiliation{%
  \institution{New York University}
  \city{New York}
  \state{New York}
  \country{USA}
}

\begin{document}

\title{ACT-Tensor: Tensor Completion Framework for Financial Dataset Imputation}

\renewcommand{\shortauthors}{Mo, Li, Zhang, Chen}

\begin{abstract}

Missing data in financial panels presents a critical obstacle, undermining asset-pricing models and reducing the effectiveness of investment strategies. Such panels are often inherently multi-dimensional, spanning firms, time, and financial variables, which adds complexity to the imputation task. Conventional imputation methods often fail by flattening the data's multidimensional structure, struggling with heterogeneous missingness patterns, or overfitting in the face of extreme data sparsity. To address these limitations, we introduce an Adaptive, Cluster-based Temporal smoothing tensor completion framework (ACT‑Tensor) tailored for severely and heterogeneously missing multi-dimensional financial data panels. ACT-Tensor incorporates two key innovations: a cluster‑based completion module that captures cross‑sectional heterogeneity by learning group-specific latent structures; and a temporal smoothing module that proactively removes short‑lived noise while preserving slow‑moving fundamental trends. Extensive experiments show that ACT‑Tensor consistently outperforms state-of-the-art benchmarks in terms of imputation accuracy across a range of missing data regimes, including extreme sparsity scenarios. To assess its practical financial utility, we evaluate the imputed data with an asset‐pricing pipeline tailored for tensor-structured financial data. Results show that ACT-Tensor not only reduces pricing errors but also significantly improves risk-adjusted returns of the constructed portfolio.  These findings confirm that our method delivers highly accurate and informative imputations, offering substantial value for financial decision-making.

\end{abstract}

\ccsdesc[500]{Applied computing~Operations research}
\ccsdesc[500]{Computing methodologies~Artificial Intelligence}

\keywords{
Asset Pricing, Missing Data Imputation, Tensor Completion, Factor Models}

\maketitle

\section{Introduction} 
Financial data—particularly firm-level characteristic data—are widely used by researchers and practitioners in empirical asset pricing to explain and predict the cross-section of expected returns \cite{fama1992cross, fama1993common, carhart1997persistence, pastor2003liquidity} and to support the construction of systematic investment strategies \cite{asness2019quality}. These data, indexed by firm, characteristic, and time, naturally form a multi-dimensional structure, which not only captures temporal dynamics but also preserves cross-sectional heterogeneity across firms. The financial utility of asset pricing models critically depends on the completeness and quality of these multi-dimensional panels, as reliable inputs are essential for effective investment decisions.

However, in practice, such fully observed firm‑level characteristic panels are not always available. For example, CRSP/Compustat data show that at any given month, more than 70\% of listed equities lack at least one characteristic, accounting for roughly half of the market’s total capitalization \cite{bryzgalova2025missing}. Moreover, this missingness is systematic: smaller, younger, or financially distressed firms are disproportionately incomplete \cite{freyberger2025missing, easton2020attrition}. When these extensive, non‑random missing data are handled with overly simplistic or inadequate methods, they embed systematic bias into asset pricing models, which leads to significant pricing errors and sub-optimal investment strategies.

Early solutions, such as cross-sectional median filling \cite{gu2020empirical, kozak2020shrinking, light2017aggregation} and discarding incomplete observations \cite{freyberger2020dissecting, lewellen2014cross,kelly2019characteristics} either attenuate the signal or induce selection bias. Nevertheless, more refined techniques offer only partial relief. The generalized method of moments (GMM) framework \cite{freyberger2025missing} relies on a missing-at-random (MAR) assumption, while 
Expectation-Maximization (EM) algorithms \cite{chen2022missing,jin2021factor,banbura2014maximum} fill the gaps under even stricter and rarely satisfied assumptions, such as joint-normality. Matrix‑based imputations \cite{bryzgalova2025missing} flatten the three‑dimensional panel into a two‑dimensional matrix, discarding the time dimension and the temporal dependence essential to financial characteristics.

More recent work preserves the panel's three-way structure by adapting tensor completion algorithms \cite{zhou2023fast,acar2011all,choi2019s3}, while several key challenges remain unresolved. First, extreme sparsity of data breeds over‑fitting, and theory shows reconstruction error grows unbounded as data coverage shrinks \cite{montanari2018spectral, mu2014square, liu2020tensor}. Second, the single latent factor structure applied to all firms ignores the firm-level heterogeneity \cite{freyberger2025missing, bryzgalova2025missing}. Most models treat the time dimension as a conventional variable \cite{acar2011all, choi2019s3}, neglecting its unique properties. Hence, they often fail to account for the non-stationarity inherent in firm characteristics.

To address the challenges posed by missing data in financial panels, we propose an Adaptive, Cluster-based Temporal smoothing tensor completion framework (ACT-Tensor) designed to handle sparse, multidimensional financial datasets. Unlike conventional methods, which either discard the rich cross‑sectional and temporal dependencies or fail to properly address extreme sparsity, ACT‑Tensor is specifically built to preserve the multidimensional structure of the data, ensuring more stable factor estimates and accurate return forecasts. In general, our main contributions are threefold.

First, we enhance tensor completion with two innovative modules: (i) a cluster‑based completion module that captures the rich cross‑sectional heterogeneity to address overfitting under extreme sparsity and avoid model bias in methods that apply a single global structure; and (ii) a temporal smoothing module that filters short‑lived noise while preserving persistent signals to ensure imputed values reflect long‑term trends and remain robust to non-stationarity. Together, they overcome the limitations of existing methods that fail to capture firm-level heterogeneity and temporal dependencies.

Second, we deploy a state-of-art asset-pricing model that directly feeds imputed tensor data into portfolio construction and return forecasting. Hence, the quality of imputed data is assessed not just by reconstruction error but by the accuracy of the resulting pricing models and the profitability of the strategies.

Third, extensive experiments demonstrate that our approach consistently not only achieves superior imputation accuracy across diverse missing-data regimes but also converts that accuracy into markedly stronger performance on asset pricing tasks, ultimately enabling more profitable, better risk-adjusted investment strategies.

\section{Related Work}

Missing data in financial asset panels is a pervasive problem that has long been under-acknowledged in asset pricing research. When extensive, non-random missing data are filled heuristically or ignored, they can embed systematic biases into asset pricing models, leading to significant pricing errors and distorted inferences that undermine investment strategies. 

\medskip
\noindent
\textbf{Early Approaches.} 
Early solutions for missing data involved filling missing values with cross-sectional medians \cite{gu2020empirical, kozak2020shrinking, light2017aggregation} or confining analysis to the minority of firms with fully observed data \cite{lewellen2014cross, freyberger2020dissecting, kelly2019characteristics}. While straightforward to implement, empirical evidence has shown that these ad-hoc strategies introduce significant bias and undermine the robustness of model estimations \cite{gu2020empirical}.

\medskip
\noindent
\textbf{Matrix Factorization Based Models.} 
To exploit the cross-sectional correlations, many approaches collapse the three-way panel into matrices and applying principal‑component analysis \cite{bai2017principal, cahan2023factor} or matrix completion methods \cite{jin2021factor, bryzgalova2025missing}. While this flatten-and-factor strategy improves upon simple heuristics, it applies a single low‑rank structure on all firms and discard both temporal dependencies \cite{lettau20243d} and firm-level heterogeneous effects \cite{acar2011all}.

\medskip
\noindent
\textbf{Estimation Based Models.}
Another line of methods integrate imputation directly into model estimation. GMM‑based methods \cite{freyberger2025missing} require a MAR assumption, where the probability that a characteristic is missing is independent of its unobserved value. Similarly, EM-based techniques \cite{chen2022missing,jin2021factor,banbura2014maximum} rely on strong distributional assumptions, such as joint normality, and impute each period in isolation, discarding valuable time-series information. Consequently, both families of methods can produce biased or unstable estimates when data are sparse.

\medskip
\noindent
\textbf{Tensor Completion Based Models.}  
A body of work preserves the panel’s three-way structure by applying low-rank tensor decompositions for imputation \cite{acar2011all,choi2019s3,zhou2023fast}. Empirically, these tensor methods outperform matrix-based approaches, especially when data are sparse \cite{zhou2023fast,han2024cp}. However, key limitations remain: (i) extreme sparsity still leads to overfitting, since learning a single global rank from sparse observations leads to reconstruction error to explode as data-density shrinks \cite{montanari2018spectral,mu2014square,liu2020tensor}; (ii) the assumption of one latent structure for all firms ignores cross-sectional heterogeneity in size, industry, and life cycle \cite{freyberger2025missing,bryzgalova2025missing}; and (iii) most tensor-completion models treat the time dimension as a conventional variable and ignore its unique properties. Even time-variant extensions \cite{sun2020low,mardani2015subspace} are computationally burdensome and untested on large financial panels with extreme sparsity.

\medskip
\noindent
\textbf{Machine Learning Based Models.} Recent studies also apply deep learning techniques, including recurrent neural networks \cite{che2018recurrent, cao2018brits} and Transformers \cite{beckmeyer2022recovering, du2023saits}, to impute missing financial data. Although these architectures can model complex nonlinearities, they treat each firm’s observations as plain vectors or sequences, overlooking the rich cross-sectional and characteristic-level interactions captured by a multidimensional tensor. Moreover, deep learning models often act as black boxes with limited interpretability and theoretical guarantees, and they demand extensive hyperparameter tuning and regularization to avoid overfitting. Consequently, purely ML-based imputations struggle to deliver consistently high accuracy without embedding domain-specific structure.

\medskip
\noindent
\textbf{Advances in Tensor and Matrix Learning.}  
Recent progress in high-dimensional matrix and tensor learning has advanced modeling, inference, and prediction for multi-way data.  
In matrix-variate settings, constrained factor models capture cross-sectional and temporal dependence \cite{chen2019constrained}, threshold structures handle regime shifts \cite{liu2022identification}, and dynamic matrix-factor approaches extend to transport and spatial-temporal systems \cite{chen2022modeling, chen2020modeling}.  
Inference for high-dimensional matrix factors \cite{chen2023statistical} and factor-augmented regression \cite{chen2024factor} further enhance interpretability and robustness.

For tensors, low-rank CP/Tucker decompositions exploit multi-modal structure for efficient estimation.  
Semiparametric tensor factor analysis via iteratively projected SVD attains rate-optimal recovery \cite{chen2024semi, chen2020semiparametric}, and distributed tensor PCA addresses heterogeneous data across sites \cite{chen2025distributed}.  
Supervised tensor methods extend to discriminant analysis and classification with CP-structured low-rank forms, including incomplete tensors \cite{chen2024high, chen2024highb}.  
Tensor structures have also been embedded into deep models such as tensor-view topological GNNs \cite{wen2024tensor}, tensor-fused graph contrastive learning \cite{wu2025tensor}, tensor-augmented Transformers \cite{kong2024teaformers}, and tensorized uncertainty quantification \cite{wu2024conditional,wu2024conditionalb}.  
Building on this literature, our work adapts tensor learning to large, sparsely observed financial panels.  
ACT-Tensor integrates low-rank estimation, distributed learning, and temporal regularization, bridging statistical tensor modeling with practical financial imputation and pricing applications.

In summary, we overcome the challenges faced by existing methods through introducing  ACT‑Tensor with two innovative modules to (i) handle extreme data sparsity without over‑fitting, (ii) capture firm‑level heterogeneity via cluster‑based completion, and (iii) accommodate temporal non‑stationarity with data‑driven smoothing. Unlike prior studies that judge success only by statistical error, we also evaluate imputations by their impact on pricing accuracy and return forecasts. The comprehensive result shows that ACT‑Tensor delivers state‑of‑the‑art recovery of missing data and significantly improves pricing accuracy, thereby demonstrating the practical value of a unified approach that rigorously connects imputation advances to asset‑pricing performance.

\section{ACT-Tensor Imputation Framework}

Missing data in firm‑characteristic panels undermines factor estimates and return forecasts in asset‑pricing models. To counter this problem and overcome the shortcomings of existing imputation methods, we introduce ACT-Tensor, which divides the problem into two innovative, complementary modules: (i) cluster-based completion (\autoref{subsec:cluster_aware}) and (ii) temporal smoothing (\autoref{subsec:temporal_smooth}). 

\subsection{Tensor Factorization and Completion}
We treat the firm-characteristic panel as a third‐order tensor $\mathcal{X}\in\mathbb{R}^{T\times N\times L}$, where \(T\) is the number of months, 
\(N\) is the number of firms, and \(L\) is the number of characteristics. To exploit the low-rank structure of $\mathcal{X}$, we adopt the CANDECOMP/PARAFAC (CP) decomposition. For a chosen rank \(R\), the completed tensor \(\hat{\mathcal{X}}\) is written in terms of three loading matrices \(U\in\mathbb{R}^{T\times R}\), \(V\in\mathbb{R}^{N\times R}\), and \(W\in\mathbb{R}^{L\times R}\) with component weights $\boldsymbol{\gamma} = (\gamma_1,\dots,\gamma_R)^\intercal$:
\begin{equation}
    \hat{\mathcal{X}} = [\![\boldsymbol{\gamma}\,;\,U,\,V,\,W\,]\!] = \sum_{r=1}^{R} \gamma_r\,\mathbf{u}_r \otimes \mathbf{v}_r \otimes \mathbf{w}_r.
\end{equation}
To remove scale ambiguity we set $\gamma_r = 1$ for $r = 1,\dots,R$ \cite{kolda2009tensor}. Each entry of $\hat{\mathcal{X}}$ is simply:
\begin{equation}
    \hat{x}_{t,n,\ell} = \sum_{r=1}^{R} \gamma_r \, U_{tr} \,V_{nr}\,W_{\ell r},
\end{equation}
where $\mathbf{u}_r \in \mathbb{R}^T$, $\mathbf{v}_r \in \mathbb{R}^N$, and $\mathbf{w}_r \in \mathbb{R}^L$ are the $r$th columns of $U$, $V$, and $W$. When the original tensor $\mathcal{X}$ contains missing values, we estimate $(U,V,W)$ by minimizing the reconstruction error over the observed set $\Omega$:
\begin{equation}
  \min_{U,V,W} \left\|\mathcal{P}_{\Omega}(\mathcal{X})-\mathcal{P}_{\Omega}\left([\![U,V,W]\!]\right)\right\|_{F}^{2}+\lambda\left(\|U\|_{F}^{2}+\|V\|_{F}^{2}+\|W\|_{F}^{2}\right),
\end{equation}
where $\|\cdot\|_F$ denotes the Frobenius norm, \(\mathcal{P}_{\Omega}\) masks unobserved entries and \(\lambda\) is an \(\ell_{2}\) regularizer that guards against over‑fitting. 

Although ACT-Tensor is fully compatible with Tucker or Tensor‐Train decompositions, we focus on CP here for its computational simplicity and its proven effectiveness in large, sparse settings \cite{acar2011all,zhou2023fast}, avoiding the dense cores or multiple‐core overhead required by alternative tensor decompositions.

\subsection{Cluster-based Completion}
\label{subsec:cluster_aware}
Standard tensor completion algorithms break down when the data are sparse due to overfitting. To stabilize the fit, we introduce a cluster-based completion strategy. We first partition firms by data density, classifying each cluster as dense or sparse (\autoref{subsubsec:firm_clustering}). Dense clusters are completed independently, whereas sparse clusters are augmented with data from their dense peers before completion (\autoref{subsubsec:cluster_imputation}). The resulting sub-tensors are then reassembled into the fully imputed panel (\autoref{subsubsec:assembly}).

\subsubsection{Clustering of Companies}
\label{subsubsec:firm_clustering}
We begin by grouping firms according to their observed-entry rate to separate the firms with sparse observations from the ones with sufficient data. 

For each firm \(n=1,\dots, N\), we group firms by their observed time–characteristic patterns \(T\times L\) as:
\begin{equation}
    \mathbf{X}_{n}=\mathcal{X}\left({:,n,:}\right)\in\mathbb{R}^{T\times L}, \quad \mathbf{v}_{n}=\operatorname{vec}\!\left(\mathbf{X}_{n}\right)\in\mathbb{R}^{TL},
\end{equation}
where $\operatorname{vec}(\,\cdot\,)$ stacks the columns of a matrix into a single vector. Then, we partition the set $\{\mathbf{v}_{n}\}_{n=1}^{N}$ into $K$ clusters via K-means:
\begin{equation}
    \min_{z\in\{1,\dots,K\}^{N}, \,\{\boldsymbol \mu_{k}\}_{k=1}^{K}} \,\sum_{n=1}^{N} \left\lVert \mathbf{v}_{n}-\boldsymbol\mu_{\,z_{n}} \right\rVert_{2}^{2},
\end{equation}
where \(z_{n}\) is firm \(n\)’s cluster label and \(\mu_{k}\in\mathbb{R}^{TL}\) is the \(k\)th centroid. Let \(\mathcal{I}_{k}=\{n:z_{n}=k\}\) be the firms in cluster \(k\), and define its observed‐entry ratio as:
\begin{equation}
  \rho_{k}
  = \frac{1}{|\mathcal{I}_{k}|\,T\,L}
    \sum_{n\in\mathcal{I}_{k}}
    \sum_{t=1}^{T}
    \sum_{\ell=1}^{L}
      \mathbf{1}\left[(t,n,\ell)\in\Omega\right],
\end{equation}
where \(\Omega\) indexes the observed entries. Then, the sub-tensor of cluster $k$ can be denoted as:
\begin{equation}
    \mathcal{X}_k
  = \mathcal{X}_{:\,,\,\mathcal{I}_k\,,\,:}
  \;\in\;\mathbb{R}^{T\times|\mathcal{I}_k|\times L}.
\end{equation}
Given a threshold \(\tau\), we call cluster \(k\) dense if \(\rho_{k}\ge\tau\) and sparse otherwise. $\tau=40\%$ is chosen to ensure a stable completion \cite{mu2014square, zhou2023fast, jain2014provable}. Thus, after density–driven clustering, we categorize the clusters into dense \(\mathcal{D}=\{d:\rho_d\ge\tau\}\) and sparse \(\mathcal{S}=\{s:\rho_s<\tau\}\).

\subsubsection{Cluster-based Completion}
\label{subsubsec:cluster_imputation}
After clustering of firms, we handle the two types of clusters accordingly. 

\medskip\noindent
\textbf{Dense Clusters.} For dense clusters, we apply tensor completion directly to their own sub-tensors. For each \(d\in\mathcal{D}\), the sub‐tensor of dense cluster $d$ is represnted as:
\begin{equation}
  \mathcal{X}_d
  = \mathcal{X}_{:\,,\,\mathcal{I}_d\,,\,:}
  \;\in\;\mathbb{R}^{T\times|\mathcal{I}_d|\times L}
\end{equation}
We then solve a rank-$R$ CP problem on its observed entries, setting $\lambda=0$ because over‑fitting is not a concern at this density:
\begin{equation}
  \min_{U,\;V_d,\;W}
  \bigl\|
    \mathcal{P}_{\Omega_d}(\mathcal{X}_d)
  - \mathcal{P}_{\Omega_d}\left([\![U,V_d,W]\!]\right)
  \bigr\|_F^2,
\end{equation}
where \(V_d\in\mathbb{R}^{|\mathcal{I}_d|\times R}\). Then, we can get the imputed sub-tensor \(\hat{\mathcal{X}}_d\). 

\medskip
\noindent
\textbf{Sparse Clusters.} 
For each sparse cluster $s$, we augment its sparse panel by building an aggregated tensor with data from all dense clusters before completing it. Specifically, we form: 
\begin{equation}
  \mathcal{X}^{\rm agg}_s
  = \mathcal{X}_{:\,,\,\mathcal{I}_s^{\rm agg}\,,\,:}
  \in \mathbb{R}^{T\times|\mathcal{I}_s^{\rm agg}|\times L}, \quad \text{with} \;
  \mathcal{I}_s^{\rm agg}
  = \mathcal{I}_s \cup \bigcup_{d\in\mathcal{D}}\mathcal{I}_d.
\end{equation}
We then solve a rank-$R$ CP problem on the observed entries of this aggregated tensor, using a $\ell_2$ penalty:
\begin{equation}\label{eq:sparse_cp}
\begin{split}
  \min_{U,\,V^{\rm agg}_s,\,W}\;
    &\left\|
      \mathcal{P}_{\Omega^{\rm agg}_s}(\mathcal{X}^{\rm agg}_s)
    - \mathcal{P}_{\Omega^{\rm agg}_s}\left([\![U,V^{\rm agg}_s,W]\!]\right)
    \right\|_F^2\\
  &\quad+
    \lambda\left(\|U\|_F^2 + \|V^{\rm agg}_s\|_F^2 + \|W\|_F^2\right).
\end{split}
\end{equation}
where \(V^{\rm agg}_s\in\mathbb{R}^{|\mathcal{I}_s^{\rm agg}|\times R}\),
and \(\Omega^{\rm agg}_s\) indexes all observed entries.
After fitting, we slice out the completed sub-tensor that belongs to the sparse cluster itself:
\begin{equation}
  \hat{\mathcal{X}}_d
  = (\hat{\mathcal{X}}^{\mathrm{agg}}_{s})_{:\,,\,\mathcal{I}_{s}\,,\,:}
  \;\in\;\mathbb{R}^{T\times|\mathcal{I}_{s}|\times L}.
\end{equation}
Repeating this for every \(s\in\mathcal{S}\) produces completed sub‐tensors
for all sparse clusters. 

\subsubsection{Assembly of Completed Sub-tensors}
\label{subsubsec:assembly}
After each cluster has been completed, we rebuild the panel by putting every firm’s imputed slice back into its original position. Let \(\hat{\mathcal{X}}_k\in\mathbb{R}^{T\times|\mathcal{I}_k|\times L}\)
be the completed sub-tensor for cluster \(k\).  Then the global tensor
\(\hat{\mathcal{X}}\in\mathbb{R}^{T\times N\times L}\)
is formed by:
\begin{equation}
    \hat{\mathcal{X}}_{:,\mathcal{I}_k,:}
  =\hat{\mathcal{X}}_k,
  \quad \text{with} \;
  k=1,\dots,K.
\end{equation}
Concatenating all \(K\) completed sub-tensors yields the fully imputed tensor denoted by $\hat{\mathcal{X}}$, which is then passed to the temporal‐smoothing module.  

\subsection{Temporal Smoothing}
\label{subsec:temporal_smooth}
While cluster‑aware completion captures high-level time patterns through its temporal factors, it can still leave short‑lived noise that obscures the true sequential dynamics. To restore underlying dynamics, we add a temporal smoothing module that filters each firm–characteristic series before the final panel is used in asset‑pricing tests. We consider three smoother options: centered moving average (\autoref{subsubsec:CMA}), exponential moving average (\autoref{subsubsec:EMA}), or Kalman filter (\autoref{subsubsec:KF}). 

\subsubsection{Centered Moving Average (CMA)} 
\label{subsubsec:CMA}
To smooth out short-lived noise without flattening medium‑term trends in the data, we apply a centered moving average to each imputed series. Specifically, for window size \(\delta\) with \(m = (\delta - 1)/2\):
\begin{equation}
    \tilde{x}_{t}^{n,\ell} =\frac{1}{\delta}\sum_{s=-m}^{m}\hat{x}_{t+s}^{n,\ell} \quad \text{with} \; t = m+1,\dots,T-m
\end{equation}
where $\hat{x}_{t+s}^{n,\ell}$ denotes the imputed data of company $n$'s characteristic $\ell$ at $t+s$. Near the ends of the series (\(t \le m\) or \(t > T-m\)), we shrink the window to include only available observations. 

\subsubsection{Exponential Moving Average (EMA)}
\label{subsubsec:EMA}
We choose exponential moving average as it reacts more rapidly to structural breaks or permanent shifts than a symmetric filter would while still effectively smoothing idiosyncratic noise. With smoothing factor \(\theta\in(0,1)\), we apply the following recursive formula:
\begin{equation}
    \tilde{x}_{1}^{n,\ell} = \hat{x}_{1}^{n,\ell}, \quad \tilde{x}_{t}^{n,\ell} = \theta\,\hat{x}_{t}^{n,\ell} + (1-\theta)\,\tilde{x}_{t-1}^{n,\ell}, \quad \text{with} \; t=2,\dots,T.
\end{equation}

\subsubsection{Kalman Filter (KF)}
\label{subsubsec:KF}
The Kalman filter offers a probabilistic way to smooth each series and quantify its remaining uncertainty. We treat the latent true value $y_{t}^{n,\ell}$ as a random walk:
\begin{equation}
  y_{t}^{n,\ell} = y_{t-1}^{n,\ell} + w_{t}, \quad \text{with} \; w_{t}\sim\mathcal{N} (0,h),  
\end{equation}
and view each imputed point as a noisy observation,
\begin{equation}
   \hat{x}_{t}^{n,\ell} = y_{t}^{n,\ell} + v_{t}, \quad \text{with} \; v_{t}\sim\mathcal{N}(0,r).
\end{equation}
where \(h\) and \(r\) controls process‐ and measurement-noise variance, respectively. Starting from a prior \((y_{1}^{n,\ell},P_{1}^{n,\ell})\), the standard predict–update recursion produces filtered means \(\bar y_{t}^{n,\ell}\) and variances \(P_{t}^{n,\ell}\). A backward smoothing pass then yields
\begin{equation}
  \tilde{x}_{t}^{n,\ell} = \mathbb{E}\left[y_{t}^{n,\ell}\mid \hat{x}_{1:T}^{n,\ell}\right],
\end{equation}
which we take as the final, temporally smoothed imputation.  

We select all hyperparameters in the temporal smoothing module by grid-search. The resulting smoothed tensor $\tilde{\mathcal{X}} \in \mathbb{R}^{T \times N \times L}$ is the input for the asset-pricing applications described in \autoref{sec:asset_pricing}.

\subsection{Imputation Accuracy Metrics}
\label{subsec:imputation_metrics}
To assess how closely each method reconstructs the held‑out characteristic values, we report four widely used error measures. Let the evaluation set contain \(M\) masked entries for evaluation with true values \(x_{m}\) and imputed values \(\tilde x_{m}\) for \(m=1,\ldots,M\):

\begin{equation}
\begin{aligned}
\text{RMSE}_{\text{imp}} &= \sqrt{\frac{1}{M}\sum_{m=1}^{M}(x_{m}-\tilde x_{m})^{2}}, \quad
\text{MAE}_{\text{imp}}  =  \frac{1}{M}\sum_{m=1}^{M}\lvert x_{m}-\tilde x_{m}\rvert, \\
\text{MAPE}_{\text{imp}} &= \frac{1}{M}\sum_{m=1}^{M}\left\lvert\frac{x_{m}-\tilde x_{m}}{x_{m}}\right\rvert, \quad \,\,\,\,\;
R^{2}_{\text{imp}} = 1 - \frac{\sum_{m}(x_{m}-\tilde x_{m})^{2}}
                              {\sum_{m}(x_{m}-\bar x)^{2}}.
\end{aligned}
\end{equation}
where \(\bar x = \tfrac{1}{M}\sum_{m} x_{m}\) is the sample mean of the true values. 

The full algorithm of ACT-Tensor is shown in \autoref{algo:ClusterTensor}.

\begin{algorithm}[t]
  \caption{ACT-Tensor Framework}
  \label{algo:ClusterTensor}
  \KwIn{Target tensor $\mathcal{X}\in\mathbb{R}^{T\times N\times L}$ to be imputed,  tensor decomposition rank $R$, number of clusters $K$, density threshold $\tau$, smoothing method $\in\{\mathrm{CMA},\mathrm{EMA},\mathrm{KF}\}$ with parameters $(\alpha,\beta)$}
  \KwOut{Smoothed imputed tensor $\tilde{\mathcal{X}} \in \mathbb{R}^{T\times N\times L}$}
  // \textbf{Clustering of Companies} \\
  Compute $\mathbf{v}_n = \mathrm{vec}(\mathcal{X}_{:\,,\,n\,,\,:})$ for $n=1\dots N$\;
  Run K–means with $K$ clusters on $\{\mathbf{v}_n\}$ to gain $\mathcal{D}$ and $\mathcal{S}$ by calculating density $\rho_k$ for each $k$\;
  // \textbf{Cluster–wise CP Completion} \\
  \For{$d \in \mathcal{D}$}{
    CP completion on sub–tensor $\mathcal{X}_d\;\to\;\hat{\mathcal{X}}_d$\;
  }
  \For{$s \in \mathcal{S}$}{
      Form aggregated tensor $\mathcal{X}^{\mathrm{agg}}_{s}$ with firms $\mathcal{I}_{s} \cup \bigcup_{d\in \mathcal{D}}\mathcal{I}_d$\;
      CP completion on aggregated sub–tensor $\mathcal{X}^{\mathrm{agg}}_{s}$\; 
      Slice rows corresponding to $\mathcal{I}_{s}$ to obtain $\hat{\mathcal{X}}_s$\;
  }
  // \textbf{Global Assembly of Imputed Sub-tensors} \\
  Initialise $\hat{\mathcal{X}} = \mathbf{0} \in \mathbb{R}^{T\times N\times L}$\;
  \For{$k=1\dots K$}{Insert $\hat{\mathcal{X}}_{k}$ into firm positions $\mathcal{I}_k$ of $\hat{\mathcal{X}}$\;}
  // \textbf{Temporal Smoothing} \\
  \For{$n=1\dots N$, $\ell=1\dots L$}{
    $\tilde{x}_{t}^{n,\ell} \leftarrow \mathrm{Smooth}\left(\hat{x}_{t}^{n,\ell};\,S,\,\alpha,\beta\right)$
  }
  \Return $\tilde{\mathcal{X}} \in \mathbb{R}^{T\times N\times L}$
\end{algorithm}

\section{Downstream Asset Pricing Evaluation}
\label{sec:asset_pricing}
This section explores whether the imputed characteristic panels capture genuine financial signals rather than just noise with an advanced asset‐pricing pipeline tailored for tensor-structured financial data \cite{lettau20243d}. We first convert the panel into investable strategies by forming value‑weighted portfolios, creating an excess return tensor (\autoref{subsec:portfolio}). We compress this tensor into a small set of latent return drivers with tensor decomposition to identify which of those factors genuinely forecast future returns (\autoref{subsec:factor_extraction}). Then, we quantify their predictive strength and rebuild a filtered excess return tensor, providing the foundation for evaluation metrics reported later (\autoref{subsec:factor_evaluation}).

\subsection{Excess Return Tensor Construction} 
\label{subsec:portfolio}
At each time spot $t$, we form value‑weighted portfolios based on the firm-characteristic panel with a double sorting scheme. Firms are first split into $P$ size buckets; within each size bucket we rank the firms on each of the remaining $(L-1)$ characteristics, creating $Q$ sub-baskets. We denote by $\{i \in \mathcal{B}_{p,q,\ell,t}\}$ the set of firms that fall into size bucket \(p\) and characteristic‑\(\ell\) rank bucket \(q\). The value-weighted average excess return of $\mathcal{B}_{p,q,\ell,t}$is computed as:
\begin{equation}
    \mathcal{R}_{p,q,\ell,t}
    = \frac{\sum_{i\in\mathcal{B}_{p,q,\ell,t}} w_{i,t}\,r_{i,t}}
           {\sum_{i\in\mathcal{B}_{p,q,\ell,t}} w_{i,t}}
      - r_{f,t},
\end{equation}
where \(w_{i,t}\) is firm \(i\)’s market capitalization, \(r_{i,t}\) is firm \(i\)’s realized return, and \(r_{f,t}\) is the risk-free rate. Stacking these returns over time and characteristics produces the portfolio‑return tensor:
\begin{equation}
    \mathcal{R}\in\mathbb{R}^{P\times Q\times (L-1)\times T}.
\end{equation}

\subsection{Tensor‑Based Factor Extraction}
\label{subsec:factor_extraction}
To uncover a parsimonious set of return drivers, we perform a rank \((k_p,k_q,k_\ell)\) partial Tucker decomposition of \(\mathcal{R}\). Since we are only interested in the factor structure of the portfolios, the time dimension is not factored. In order to find the core tensor $\mathcal{F} \in \mathbb{R}^{k_p \times k_q \times k_\ell \times T}$ and loading matrices ${U}\in\mathbb{R}^{P\times k_p}$, ${V}\in\mathbb{R}^{Q\times k_q}$, and ${W}\in\mathbb{R}^{(L-1)\times k_\ell}$, we solve the minimization problem of the approximation error:
\begin{equation}
\label{eq:partial_tucker}
    \operatorname*{argmin} \|\mathcal{R} - {\mathcal{F}} \times_2 U \times_3 V \times_4 W\|_F.
\end{equation}
As \Eqref{eq:partial_tucker} does not have a closed-form solution, we apply Higher-Order SVD \cite{de2000multilinear} to approximate $U$, $V$, $W$ and get $\hat{U}\in\mathbb{R}^{P\times k_p}$, $\hat{V}\in\mathbb{R}^{Q\times k_q}$, and  $\hat{W}\in\mathbb{R}^{(L-1)\times k_\ell}$. After obtaining the loading matrices, we project back to get \(\hat{\mathcal{F}}\):
\begin{equation}
    \hat{\mathcal{F}} = \mathcal{R} \times_2 \hat{U}^\intercal \times_3 \hat{V}^\intercal \times_4 \hat{W}^\intercal,
\end{equation}
where $\hat{\mathcal{F}} \in \mathbb{R}^{k_p \times k_q \times k_\ell \times T}$. By stacking the \(k=k_p\,k_q\,k_\ell\) time‐mode slices of \(\hat{\mathcal{F}}\) at each \(t\) into $\mathbf{f}_{t}\in\mathbb{R}^{k}$, we obtain a compact latent state that is a concise summary of the key drivers of cross-sectional returns at $t$. This latent state serves as the input for the return-prediction regression described below.

\subsection{Factor-Based Return Prediction}
\label{subsec:factor_evaluation}
To evaluate which latent factors truly forecast future returns, for every portfolio \(n=1,\dots,N\) portfolios, we run the following time-series regression: 
\begin{equation}
  {r}_{n, t+1}= \alpha_n + \boldsymbol{\beta}_n^{\intercal}\mathbf{f}_t^{(\mathcal{M})} + \varepsilon_{n, t+1},
\end{equation}
where $r_{n, t+1}$ is the excess return of portfolio $n$ next period, $\mathbf{f}_t^{(\mathcal{M})}$ is a candidate subset of the $k$ latent factors, and \(\alpha_n\) is the pricing‑error intercept. We choose the subset $\mathcal{M}$ with a forward stepwise search that maximize the pseudo cross-sectional $R^2$:
\begin{equation}
  R^2_{\text{xs}} = 1 - \frac{\frac{1}{N}\sum_{n=1}^N \alpha_n^2}
           {\displaystyle\operatorname{Var}_{\text{xs}}(\bar R_n)},
\end{equation}
adding one factor at a time until the target model size $|\mathcal{M}|$ is reached. This yields the most informative, yet parsimonious, factor set for prediction.
With the selected factor set and the estimated loadings, we can compute the model‑implied forecasts:
\begin{equation}
\label{eq:ts_reg}
    \hat r_{n,t+1} = \hat\alpha_{n}+\hat{\boldsymbol{\beta}}_n^{\intercal} \mathbf f_{t}^{(\mathcal M)},
\end{equation}  
which serve as the inputs for the evaluation metrics defined in \autoref{subsec:asset_pricing_metrics}.

\subsection{Asset Pricing Evaluation Metrics}
\label{subsec:asset_pricing_metrics}
The factor‑filtered return tensor is assessed along two complementary dimensions: pricing accuracy and predictive power.

\subsubsection{Pricing Accuracy.} For each portfolio $n$, the return-forecasting regression (\Eqref{eq:ts_reg}) yields a pricing error intercept $\alpha_n$.
We summarize these errors with root‑mean‑squared and mean‑absolute measure \cite{lettau20243d}:
\begin{equation}
    {\text{RMSE}_{\alpha}
       = \sqrt{\frac{1}{N}\sum_{n=1}^{N}\hat{\alpha}_{n}^{2}}}, \qquad {\text{MAE}_{\alpha}
       = \frac{1}{N}\sum_{n=1}^{N}\left|\hat{\alpha}_{n}\right|}.
\end{equation}
Lower values indicate that the selected factors account for a larger share of the cross‑sectional return variation.

\subsubsection{Predictive Power.} 
We quantify how well the model ranks portfolios according to next–period returns and whether that ranking can be converted into a profitable strategy. 

Following \cite{gu2020empirical}, we measure the gap between predicted and realized cross‑sectional rankings with the mean‑absolute rank error:
\begin{equation}
    \text{MAE‑Rank} = \frac{1}{T} \sum_{t=1}^{T} \frac{1}{N} \sum_{n=1}^{N} \left|\operatorname{rank}\left(r_{n,t+1}\right) - \operatorname{rank}\left(\hat{r}_{n,t+1}\right) \right|.
\end{equation}
A smaller MAE-Rank value indicates fewer mis‑ordered portfolios. 

The Information Coefficient (IC) \cite{grinold2000active} captures the linear association between predictions and outcomes:
\begin{equation}
    \text{IC} = \frac{1}{T} \sum_{t=1}^{T} \operatorname{corr} \left(\hat{\mathbf r}_{t+1},\,\mathbf r_{t+1}\right),
\end{equation}
where the correlation is computed across portfolios for each month. Higher IC signals stronger predictive alignment.

To turn ranking skill into investable profits, each month we form a Top-minus-Bottom (T-B) portfolio by longing the decile of firms with the highest predicted excess returns $\hat{r}_{n, t+1}$ and shorting the decile with the lowest \cite{fama1993common,carhart1997persistence}. The resulting excess return is:
\begin{equation}
    r^{\text{T-B}}_{t+1} = \frac{1}{N_{\text{top}}}\sum_{i\in\text{Top}} r_{i,t+1} - \frac{1}{N_{\text{bot}}}\sum_{i\in\text{Bottom}} r_{i,t+1}.
\end{equation}
where $N_{\text{top}}$ and $N_{\text{bot}}$ are the numbers of firms in the two deciles. We summarize the strategy’s risk‑adjusted performance with its annualized Sharpe ratio (T-B Sharpe):
\begin{equation}
    \text{Sharpe}_{\text{T-B}} = \sqrt{12} \cdot \frac{\bar{r}^{\text{T-B}}}{\sigma^{\text{T-B}}}.
\end{equation}
where \(\bar{r}^{\text{T-B}}\) and \(\sigma^{\text{T-B}}\) are the monthly mean and standard deviation of the T-B return series. A higher Sharpe ratio indicates that the model’s forecasts can be monetized effectively.

\section{Experiment}
\label{sec:experiment}
To evaluate the effectiveness of our imputation framework, we adopt a two‑step evaluation that assesses both the statistical accuracy of the data imputation and the financial utility that the imputed data provides in asset‑pricing applications.

First, we measure ACT-Tensor's imputation accuracy under three missing regimes. We begin with an overall panel imputation to compare ACT-Tensor against established benchmarks. Next, we zoom in on the most sparsely observed clusters to stress-test the framework's robustness under extreme sparsity conditions. These tests are supplemented by a series of ablation studies that evaluate modules' individual and joint contributions.

Second, we apply the introduced asset-pricing pipeline to assess whether the imputed data contains useful financial signals. In this step, the completed panels are passed through our evaluation pipeline to construct sorted portfolios, extract latent pricing factors, and test how well these factors predict future returns and support risk-adjusted investment strategies.

\begin{table}[H]
\centering
\caption{Setting of Parameters} 
\label{parameter_setting}
\begin{tabular}{c|c}
\toprule
Parameter Name & Parameter Value \\
\midrule
Rank $R$ & 40 \\
Number of clusters $K$ & 10 \\
Observed-entry ratio threshold $\tau$ & 40\% \\
CMA window length $\delta$ & 5 \\
EMA smoothing factor $\theta$ & 0.5 \\
Number of chosen factors $|\mathcal{M}|$ & 6 \\
Number of baskets in portfolio $P=Q$ & 20 \\
Mode-ranks $k_c$, $k_p$, $k_q$ & (5,5,5) \\
Regularization term $\lambda$ & 0 \\
\bottomrule
\end{tabular}
\end{table}

\begin{table*}[!h]
\centering
\caption{Imputation Experiment Results} 
\label{tab:imputation-errors}
\resizebox{\linewidth}{!}{%
\begin{tabular}{l|cccc|cccc|cccc}
\toprule
\multicolumn{13}{c}{\textbf{Panel A: Overall Imputation Errors}} \\
\midrule
 & \multicolumn{4}{c}{Out-of-Sample MAR} & \multicolumn{4}{c}{Out-of-Sample Block} & \multicolumn{4}{c}{Out-of-Sample Logit} \\
 \cmidrule(lr){2-5} \cmidrule(lr){6-9} \cmidrule(lr){10-13}
 \textbf{Models} & $\text{RMSE}_{\text{imp}}$ & $\text{MAE}_{\text{imp}}$ & $\text{MAPE}_{\text{imp}}$ & $R^2_{\text{imp}}$ & $\text{RMSE}_{\text{imp}}$ & $\text{MAE}_{\text{imp}}$ & $\text{MAPE}_{\text{imp}}$ & $R^2_{\text{imp}}$ & $\text{RMSE}_{\text{imp}}$ & $\text{MAE}_{\text{imp}}$ & $\text{MAPE}_{\text{imp}}$ & $R^2_{\text{imp}}$  \\
\midrule

\textbf{Cross-sectional Median} & 0.2841 & 0.2451 & \text{-} & 0.0000 & 0.2913 & 0.2532 & \text{-} & 0.0000 & 0.3135 & 0.2766 & \text{-} & 0.0000 \\

\textbf{Global BF-XS} & 0.1381 & \cellcolor{GreenDark} 0.0725 & \cellcolor{GreenMid} 1.0203 & \cellcolor{GreenDark} 0.7815 & 0.1760 & \cellcolor{GreenLight} 0.1074 & 1.5371 & 0.6027 & 0.1933 & \cellcolor{GreenLight} 0.1191 & 1.3766 & 0.4409 \\

\textbf{Local B-XS} & 0.1476 & 0.0787 & 1.1575 & \cellcolor{GreenLight} 0.7644 & 0.1838 & 0.1112 & 1.6367 & 0.6094 & 0.1966 & \cellcolor{GreenMid} 0.1181 & 1.3860 & 0.4588 \\

\textbf{ACT-Tensor w/ EMA} & \cellcolor{GreenLight} 0.1351 & \cellcolor{GreenLight} 0.0781 & 1.0963 & 0.7618 & \cellcolor{GreenLight} 0.1623 & \cellcolor{GreenMid} 0.1051 & \cellcolor{GreenLight} 1.4355 & \cellcolor{GreenLight} 0.6999 & \cellcolor{GreenLight} 0.1855 & 0.1193 & \cellcolor{GreenLight} 1.2926 & \cellcolor{GreenLight} 0.8524 \\

\textbf{ACT-Tensor w/ KF} & \cellcolor{GreenMid} 0.1344 & 0.0847 & \cellcolor{GreenDark} 0.9785 & 0.7603 & \cellcolor{GreenMid} 0.1606 & 0.1083 & \cellcolor{GreenDark} 1.2790 & \cellcolor{GreenMid} 0.7002 & \cellcolor{GreenMid} 0.1833 & 0.1219 & \cellcolor{GreenDark} 1.1834 & \cellcolor{GreenMid} 0.8578 \\

\textbf{ACT-Tensor w/ CMA} & \cellcolor{GreenDark} 0.1321 & \cellcolor{GreenMid} 0.0753 & \cellcolor{GreenLight} 1.0522 & \cellcolor{GreenMid} 0.7748 & \cellcolor{GreenDark} 0.1599 & \cellcolor{GreenDark} 0.1032 & \cellcolor{GreenMid} 1.3941 & \cellcolor{GreenDark} 0.7047 & \cellcolor{GreenDark} 0.1803 & \cellcolor{GreenDark} 0.1134 & \cellcolor{GreenMid} 1.2489 & \cellcolor{GreenDark} 0.8796 \\

\midrule

\multicolumn{13}{c}{\textbf{Panel B: Imputation Errors of Sparse Clusters}} \\
\midrule
\textbf{Cross-sectional Median} & 0.2668 & 0.2255 & \text{-} & 0.0000 & 0.2803 & 0.2394 & \text{-} & 0.0000 & 0.2901 & 0.2501 & \text{-} & 0.0000 \\
\textbf{Global BF-XS} & 0.2171 & 0.1558 & 2.2705 & 0.3871 & 0.2263 & 0.1638 & 2.1400 & 0.3648 & 0.2327 & 0.1697 & 1.9351 & 0.2271 \\
\textbf{Local B-XS} & 0.2140 & 0.1539 & 2.4546 & 0.4021 & 0.2235 & 0.1620 & 2.1884 & 0.3793 & 0.2314 & 0.1675 & 1.9883 & 0.2697 \\

\textbf{ACT-Tensor w/ EMA} & \cellcolor{GreenLight} 0.1375 & \cellcolor{GreenMid} 0.0834 & \cellcolor{GreenLight} 1.2892 & \cellcolor{GreenLight} 0.6816 & \cellcolor{GreenLight} 0.1821  & \cellcolor{GreenMid} 0.1250 & \cellcolor{GreenLight} 1.7582 & \cellcolor{GreenLight} 0.5462 & \cellcolor{GreenLight} 0.1896 & \cellcolor{GreenMid} 0.1272 & \cellcolor{GreenLight} 1.3441 & \cellcolor{GreenLight} 0.4409 \\

\textbf{ACT-Tensor w/ KF} & \cellcolor{GreenMid} 0.1368 & \cellcolor{GreenLight} 0.0884 & \cellcolor{GreenDark} 1.0980 & \cellcolor{GreenMid} 0.6854 & \cellcolor{GreenDark} 0.1799 & \cellcolor{GreenLight} 0.1262 & \cellcolor{GreenDark} 1.5746 & \cellcolor{GreenMid} 0.5501 & \cellcolor{GreenMid} 0.1868 & \cellcolor{GreenLight} 0.1278 & \cellcolor{GreenDark} 1.2357 & \cellcolor{GreenDark} 0.4612 \\

\textbf{ACT-Tensor w/ CMA} & \cellcolor{GreenDark} 0.1343 & \cellcolor{GreenDark} 0.0803 & \cellcolor{GreenMid} 1.2187 & \cellcolor{GreenDark} 0.7017 & \cellcolor{GreenMid} 0.1805 & \cellcolor{GreenDark} 0.1231 & \cellcolor{GreenMid} 1.7260 & \cellcolor{GreenDark} 0.5507 & \cellcolor{GreenDark} 0.1856 & \cellcolor{GreenDark} 0.1221 & \cellcolor{GreenMid} 1.2912 & \cellcolor{GreenMid} 0.4455 \\

\midrule

\multicolumn{13}{c}{\textbf{Panel C: Ablation Study}} \\
\midrule
\textbf{CP Completion} & 0.1511 & 0.0986 & 1.2894 & 0.6961 & 0.1695 & 0.1128 & 1.5100 & \cellcolor{GreenLight} 0.6724 & 0.1980 & 0.1338 & 1.4922 & \cellcolor{GreenLight} 0.7646 \\

\textbf{CP Completion w/ Clustering} & \cellcolor{GreenLight} 0.1431 & \cellcolor{GreenLight} 0.0922 & \cellcolor{GreenLight} 1.2601 & \cellcolor{GreenLight} 0.7292 & \cellcolor{GreenMid} 0.1632 & \cellcolor{GreenMid} 0.1070 & \cellcolor{GreenLight} 1.4553 & \cellcolor{GreenMid} 0.6958 & \cellcolor{GreenLight} 0.1938 & \cellcolor{GreenLight} 0.1293 & \cellcolor{GreenLight} 1.4448 & \cellcolor{GreenMid} 0.8016 \\

\textbf{CP Completion w/ CMA} & \cellcolor{GreenMid} 0.1351 & \cellcolor{GreenMid} 0.0782 & \cellcolor{GreenMid} 1.0689 & \cellcolor{GreenMid} 0.7585 & \cellcolor{GreenLight} 0.1655 & \cellcolor{GreenLight} 0.1085 & \cellcolor{GreenMid} 1.4515 & 0.6511 & \cellcolor{GreenMid} 0.1817 & \cellcolor{GreenMid} 0.1160 & \cellcolor{GreenMid} 1.2749 & 0.5637  \\

\textbf{ACT-Tensor w/ CMA} & \cellcolor{GreenDark} 0.1321 & \cellcolor{GreenDark} 0.0753 & \cellcolor{GreenDark} 1.0522 & \cellcolor{GreenDark} 0.7748 & \cellcolor{GreenDark} 0.1599 & \cellcolor{GreenDark} 0.1032 & \cellcolor{GreenDark} 1.3941 & \cellcolor{GreenDark} 0.7047 & \cellcolor{GreenDark} 0.1803 & \cellcolor{GreenDark} 0.1134 & \cellcolor{GreenDark} 1.2489 & \cellcolor{GreenDark} 0.8796 \\

\bottomrule
\end{tabular}%
}
\begin{flushleft}
    \footnotesize
    \textcolor{GreenDark}{\rule{1.2em}{1.2em}}\ Best performance\quad
    \textcolor{GreenMid}{\rule{1.2em}{1.2em}}\ 2\textsuperscript{nd} best\quad
    \textcolor{GreenLight}{\rule{1.2em}{1.2em}}\ 3\textsuperscript{rd} best
\end{flushleft}
\end{table*}

\subsection{Experiment Settings}

\noindent
\textbf{Dataset.} Our empirical analysis draws on the CRSP/Compustat monthly panel for 22,630 U.S. common stocks from January 2016 through December 2020. We apply standard filters to drop delisted stocks, extreme outliers, and non-exchange listings. For each firm's monthly observation, we retain 45 widely used characteristics spanning fundamentals, momentum, sentiment, and trading frictions. To ensure all characteristics are on a comparable, unit‑free scale, we cross‑sectionally rank each characteristic every month, recenter the ranks to zero, and linearly rescale them to the interval $[-0.5, 0.5]$. Roughly 83\% of the firm-characteristic entries are missing, highlighting the critical importance of a robust imputation strategy.

\medskip
\noindent
\textbf{Benchmarks.} We benchmark ACT-Tensor against both ad-hoc and state-of-the-art methods: (i) Cross-Sectional Median fills each missing data with the cross‐sectional median of that characteristic; (ii) Global BF-XS \cite{bryzgalova2025missing} applies bidirectional temporal interpolation followed by a multi-factor ridge regression; and (iii) Local B-XS \cite{bryzgalova2025missing} uses only backward filling and a multi‐factor ridge regression with a rolling window for factor estimation.

\medskip
\noindent
\textbf{Missing Regimes.} Since naturally missing entries lack ground truth values, we assess imputation accuracy by artificially introducing calibrated missingness: 10\% of the fully observed cells are masked to create a held-out testset, which we then evaluate under three masking regimes: (i) \textbf{Missing-at-random (MAR)}, where data is masked completely at random; (ii) \textbf{Block Missingness (Block)}, which randomly mask data in contiguous one-year blocks where roughly 40\% of these blocks are placed at the start of the sample, following \cite{bryzgalova2025missing}; and (iii) \textbf{Logistic Missingness (Logit)}, where a two‑stage logistic model first predicts whether a data begins with an initial gap and then assigns month‑by‑month missing probabilities conditional on past observations, producing heterogeneous, firm‑specific patterns \cite{bryzgalova2025missing}.

\medskip
\noindent
\textbf{Evaluation Metrics.} To measure imputation accuracy, we evaluate the performance of each model exclusively on the manually masked data. All metrics are calculated by comparing the imputed value against the originally known value of each masked entry. The evaluation metrics used for asset pricing and imputation accuracy evaluation are described in \autoref{subsec:imputation_metrics} and \autoref{subsec:asset_pricing_metrics}, respectively. 

For reference, the setting of parameters used in imputation and asset pricing experiments is summarized in \autoref{parameter_setting}.

\subsection{Overall Imputation Performance} 
Across all missing regimes, ACT-Tensor delivers the most accurate imputations. The results in Panel A of \autoref{tab:imputation-errors} show that ACT-Tensor consistently outperforms benchmarks under all missing regimes.  

\medskip
\noindent
\textbf{Large Gains under Structured Missingness.} ACT-Tensor's advantage is most significant when missingness follows a clear structure. In the Block regime—where entire one‑year windows disappear—our model boosts \(R^{2}_{\text{imp}}\) by 17\% over the strongest baseline, and in the firm‑dependent Logit pattern it nearly
doubles $R^2_{\text{imp}}$ from 46\% to 88\%. These gains stem from ACT–Tensor’s ability to preserve the panel's tensor structure, allowing it to recover entire blocks of data by exploiting cross‑firm co‑movements that matrix methods discard when they flatten the panel.

\medskip
\noindent
\textbf{Robustness Under Random Missingness.} While tensor completion methods are most effective when clear structures exist in the dataset, ACT-Tensor still demonstrates robust performance even when these patterns are disrupted. In the MAR scenario, which breaks up large-scale correlations, ACT-Tensor's performance margin narrows, yet it still outperforms the matrix-based methods on key error metrics, achieving an RMSE improvement of over 4.3\%.

\begin{table*}[!t]
\centering
\caption{Asset Pricing Experiment Results}
\label{downstream_task}
\resizebox{\textwidth}{!}{%
\begin{tabular}{l|ccccc|ccccc}
\toprule
 & \multicolumn{5}{c|}{Out-of-Sample Block} & \multicolumn{5}{c}{Out-of-Sample Logit} \\
\cmidrule(lr){2-6} \cmidrule(l){7-11}
\textbf{Models} & RMSE$_\alpha$ & MAE$_\alpha$ & MAE–Rank & IC & T–B Sharpe & RMSE$_\alpha$ & MAE$_\alpha$ & MAE–Rank & IC & T–B Sharpe \\
\midrule
\textbf{Global BF-XS} & 0.3975 & 0.0210 & 4942.45 & 0.1049 & 0.7293 & 0.0674 & 0.0226 & 4936.67 & 0.0761 & 0.5522  \\
\textbf{Local B-XS} & 0.1277 & 0.0270 & 4848.09 & 0.1343 & 0.7071 & 0.0609 & 0.0169 & 4898.52 & 0.0753 & 0.6353 \\
\textbf{ACT-Tensor w/ EMA} & \cellcolor{GreenLight}0.0215 & \cellcolor{GreenLight}0.0115 & \cellcolor{GreenLight}4812.26 & \cellcolor{GreenLight}0.2595 & \cellcolor{GreenLight}1.1282 & \cellcolor{GreenLight}0.0223 & \cellcolor{GreenLight}0.0118 & \cellcolor{GreenMid}4839.34 & \cellcolor{GreenMid}0.2686 & \cellcolor{GreenDark}1.0901 \\
\textbf{ACT-Tensor w/ KF} & \cellcolor{GreenMid}0.0177 & \cellcolor{GreenDark}0.0101 & \cellcolor{GreenMid}4796.51 & \cellcolor{GreenDark}0.2871 & \cellcolor{GreenDark}1.3206 & \cellcolor{GreenMid}0.0182 & \cellcolor{GreenDark}0.0105 & \cellcolor{GreenDark}4782.14 & \cellcolor{GreenDark}0.2713 & \cellcolor{GreenMid}1.0887 \\
\textbf{ACT-Tensor w/ CMA} & \cellcolor{GreenDark}0.0172 & \cellcolor{GreenMid}0.0104 & \cellcolor{GreenDark}4770.22 & \cellcolor{GreenMid}0.2850 & \cellcolor{GreenMid}1.1683 & \cellcolor{GreenDark}0.0176 & \cellcolor{GreenMid}0.0110 & \cellcolor{GreenLight}4862.01 & \cellcolor{GreenLight}0.2637 & \cellcolor{GreenLight}1.0873  
\\

\bottomrule
\end{tabular}%
}
\begin{flushleft}
    \footnotesize
    \textcolor{GreenDark}{\rule{1.2em}{1.2em}}\ Best performance\quad
    \textcolor{GreenMid}{\rule{1.2em}{1.2em}}\ 2\textsuperscript{nd} best\quad
    \textcolor{GreenLight}{\rule{1.2em}{1.2em}}\ 3\textsuperscript{rd} best
\end{flushleft}
\end{table*}

\subsection{Sparse-Cluster Stress Test}

\noindent
\textbf{Imputation Stability under Extreme Sparsity.} ACT-Tensor demonstrates exceptional stability under extreme sparsity conditions. When zooming in from the overall panel to its most challenging portion with extreme sparsity, we observe that over 80\% of companies in the dataset have fewer than 10\% of their data entries. In this demanding setting, as shown in panel B of \autoref{tab:imputation-errors}: under the MAR regime, ACT-Tensor improves $R^2_{\text{imp}}$ by an exceptional 74.5\% over the best benchmark. It achieves similarly large improvements in the Block and Logit scenarios, cutting the RMSE by approximately 20\% and increasing R² by at least 45\%, confirming its stability where other methods fail. The designed architecture of ACT-Tensor is uniquely suited for extreme sparsity, as it preserves the panel's multi-way structure, allows sparse clusters to borrow statistical strength from dense ones, and applies temporal smoothing to ensure robust outputs.

\subsection{Ablation Study}
Our ablation study demonstrates that the best-performing instantiation of ACT-Tensor adopts the CMA configuration. Furthermore, we isolate and assess the contributions of the two key modules, cluster-based completion and temporal smoothing, both individually and in combination, using vanilla CP completion as the baseline. The results are as follows:

\medskip
\noindent
\textbf{Each Module Excels in Distinct Regimes.} We isolate the cluster-based completion and temporal smoothing modules to assess their independent effectiveness. In our experiments, we compare each module against the vanilla CP completion with the same imputation settings. As shown in Panel C of Table \ref{tab:imputation-errors}, both modules improve upon the standard CP completion but excel under different missingness regimes. Temporal smoothing performs best in the MAR regime, where it leverages time‑series continuity to handle sporadic noise, while cluster-based completion delivers the largest improvements in the Block and Logit settings by capturing cross‑sectional heterogeneity.

\medskip
\noindent
\textbf{Modules' Effects Are Cumulative.} We also tested the combined effects of the two modules under different missingness regimes, and the results confirm that their synergy drives the model’s superior performance. Temporal smoothing module excels when missingness is random, leveraging time‑series continuity to handle sporadic noise. However, in cases of block‑like or firm‑specific missingness, smoothing alone can obscure important cross‑sectional differences. In these scenarios, the cluster-based completion module first groups firms with similar data density, imputing within each cluster and preserving latent patterns. ACT‑Tensor applies these steps in the following order: it first imputes within density‑matched clusters, and then applies temporal smoothing to remove short‑lived noise. This sequence ensures both stable cross‑sectional structure and robust time‑series trends, delivering the highest accuracy across all missingness regimes.

\medskip
\noindent
\textbf{CMA Outperforms Other Smoothers.} In our experiments, we tested three smoothing filters, Centered Moving Average (CMA), Exponential Moving Average (EMA), and Kalman Filtering, to evaluate their effectiveness in imputation. Among these, CMA consistently delivered the best results. Its symmetric, fixed-width window effectively removes short-lived noise while preserving slow-moving fundamental trends. In contrast, both EMA and Kalman filtering left more high-frequency noise in the imputed data, reducing their overall effectiveness. These results suggest that proactive noise suppression, rather than adaptive trend tracking, is crucial for accurate imputation, particularly in sparse financial panels where short-term fluctuations can distort underlying signals.

\subsection{Regularization‑Free Stability Test}
\begin{figure}[H]
  \centering
  \begin{subfigure}[t]{0.48\columnwidth}
    \includegraphics[width=\linewidth]{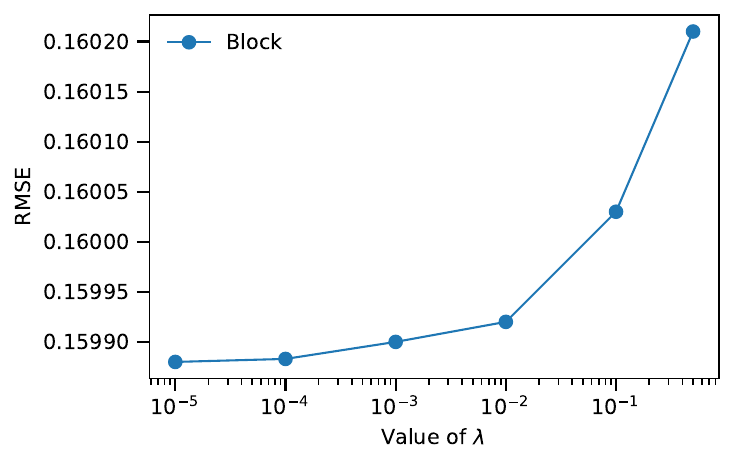}
    \caption{Block missingness}
    \label{fig:Block}
  \end{subfigure}\hfill
  \begin{subfigure}[t]{0.48\columnwidth}
    \includegraphics[width=\linewidth]{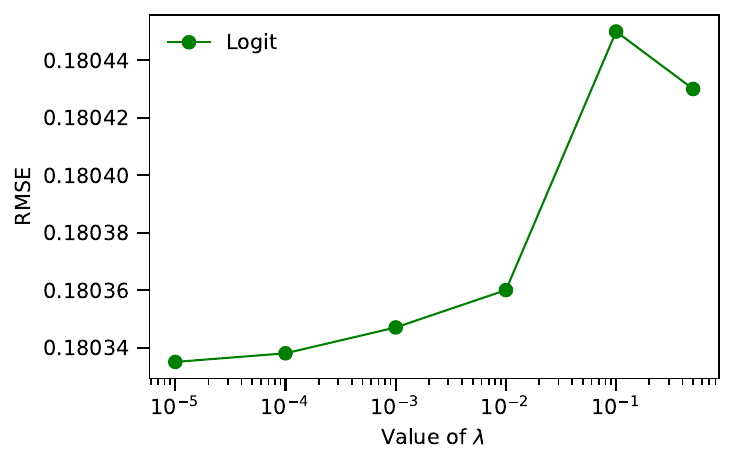}
    \caption{Logistic missingness}
    \label{fig:Logit}
  \end{subfigure}
  \caption{RMSE Sensitivity under (a) Block and (b) Logistic Missing Regimes Against the Regularization Coefficient $\lambda$.}
  \label{fig:regularization_test}
\end{figure}
We initially included an $\ell_2$ penalty in our model as a standard precaution to guard against potential overfitting, particularly when imputing the sparse clusters shown in \Eqref{eq:sparse_cp}. To determine if this regularization was truly necessary, we conducted an experiment to test its impact on performance. We systematically varied the regularization weight, $\lambda$ across a wide range from very small $10^{-5}$ to moderate ($0.5$) and recorded the out-of-sample RMSE under the Block and Logit missingness regimes.

\medskip
\noindent
\textbf{Intrinsic Stability without Need of Regularization.} The results, plotted in \autoref{fig:regularization_test}, were unambiguous: the RMSE curves remained essentially flat across the entire range of $\lambda$ values. Any change in RMSE was imperceptible, at most a few $10^{-4}$. This outcome strongly indicates that the cluster-based completion and temporal smoothing modules inherent to ACT-Tensor already provide sufficient regularization, yielding a well-conditioned optimization problem on their own. Adding ridge regularization does not improve stability or reduce error; if anything, it introduces an unnecessary bias. Based on this, we set $\lambda=0$ for all reported experiments, which has the practical benefit of simplifying the framework without the need to introduce regularization.

\subsection{Asset Pricing Performance Analysis}

\noindent
\textbf{Superior Performance on Asset Pricing Tasks.} ACT-Tensor’s imputed panels translate into markedly superior asset pricing outcomes across both missing-data regimes. As shown in \autoref{downstream_task}, our method consistently achieves the lowest pricing errors, indicating a far cleaner capture of the return–characteristic relationship. More importantly, ACT-Tensor dominates all predictive-power metrics that directly relate to financial gains. For instance, its IC more than doubles that of the next-best method, and it delivers T-B Sharpe ratio roughly twice those of conventional imputations. It also attains the lowest MAE–Rank, preserving the correct cross-sectional return ordering more faithfully than competing approaches. 

\medskip
\noindent
\textbf{Imputation Accuracy Drives Profitable Forecasts.} ACT-Tensor’s edge in both imputation and asset-pricing tests means that signals derived from its imputed data are not only statistically accurate but also highly actionable for investors, yielding markedly better risk-adjusted returns. This advantage stems from ACT-Tensor's ability to preserve the cross-sectional structure and persistent time-series information, so the imputed characteristics reflect the market’s true underlying patterns and support more reliable forecasts and trading strategies.

\section{Conclusion}
In this paper, we propose ACT-Tensor, a robust and flexible tensor completion framework that tackles the pervasive, heterogeneous missing‑data problem in asset pricing models. Our approach successfully overcomes the key limitations of conventional methods by preserving the dataset's multidimensional structure while remaining robust to the extreme and heterogeneous patterns of missingness common in financial data. The strength of ACT-Tensor lies in its design of two innovative modules: cluster-based completion, which captures the heterogeneous cross-sectional patterns, and temporal smoothing, which filters out short‑lived noise while preserving slow‑moving fundamentals. Our extensive experiments demonstrated that this architecture delivers statistically superior imputations, with particularly strong performance in the most challenging sparse-cluster stress test. Crucially, we showed that this statistical accuracy translates directly into financial utility. Data imputed by ACT-Tensor led to the construction of accurate asset-pricing models and investment strategies that deliver higher risk-adjusted returns. This confirms that our framework not only fills in missing values with high precision but also preserves the essential financial signals required for effective quantitative analysis and decision-making.

\begin{acks}
Elynn Chen's research is supported in part by the NSF Award 2412577.
\end{acks}

\bibliography{ref}


\begin{thebibliography}{53}


\ifx \showCODEN    \undefined \def \showCODEN     #1{\unskip}     \fi
\ifx \showISBNx    \undefined \def \showISBNx     #1{\unskip}     \fi
\ifx \showISBNxiii \undefined \def \showISBNxiii  #1{\unskip}     \fi
\ifx \showISSN     \undefined \def \showISSN      #1{\unskip}     \fi
\ifx \showLCCN     \undefined \def \showLCCN      #1{\unskip}     \fi
\ifx \shownote     \undefined \def \shownote      #1{#1}          \fi
\ifx \showarticletitle \undefined \def \showarticletitle #1{#1}   \fi
\ifx \showURL      \undefined \def \showURL       {\relax}        \fi
\providecommand\bibfield[2]{#2}
\providecommand\bibinfo[2]{#2}
\providecommand\natexlab[1]{#1}
\providecommand\showeprint[2][]{arXiv:#2}

\bibitem[Acar et~al\mbox{.}(2011)]%
        {acar2011all}
\bibfield{author}{\bibinfo{person}{Evrim Acar}, \bibinfo{person}{Tamara~G Kolda}, {and} \bibinfo{person}{Daniel~M Dunlavy}.} \bibinfo{year}{2011}\natexlab{}.
\newblock \showarticletitle{All-at-once optimization for coupled matrix and tensor factorizations}.
\newblock \bibinfo{journal}{\emph{arXiv preprint arXiv:1105.3422}} (\bibinfo{year}{2011}).
\newblock


\bibitem[Asness et~al\mbox{.}(2019)]%
        {asness2019quality}
\bibfield{author}{\bibinfo{person}{Clifford~S Asness}, \bibinfo{person}{Andrea Frazzini}, {and} \bibinfo{person}{Lasse~Heje Pedersen}.} \bibinfo{year}{2019}\natexlab{}.
\newblock \showarticletitle{Quality minus junk}.
\newblock \bibinfo{journal}{\emph{Review of Accounting studies}} \bibinfo{volume}{24}, \bibinfo{number}{1} (\bibinfo{year}{2019}), \bibinfo{pages}{34--112}.
\newblock


\bibitem[Bai and Ng(2017)]%
        {bai2017principal}
\bibfield{author}{\bibinfo{person}{Jushan Bai} {and} \bibinfo{person}{Serena Ng}.} \bibinfo{year}{2017}\natexlab{}.
\newblock \showarticletitle{Principal components and regularized estimation of factor models}.
\newblock \bibinfo{journal}{\emph{arXiv preprint arXiv:1708.08137}} (\bibinfo{year}{2017}).
\newblock


\bibitem[Ba{\'n}bura and Modugno(2014)]%
        {banbura2014maximum}
\bibfield{author}{\bibinfo{person}{Marta Ba{\'n}bura} {and} \bibinfo{person}{Michele Modugno}.} \bibinfo{year}{2014}\natexlab{}.
\newblock \showarticletitle{Maximum likelihood estimation of factor models on datasets with arbitrary pattern of missing data}.
\newblock \bibinfo{journal}{\emph{Journal of applied econometrics}} \bibinfo{volume}{29}, \bibinfo{number}{1} (\bibinfo{year}{2014}), \bibinfo{pages}{133--160}.
\newblock


\bibitem[Beckmeyer and Wiedemann(2022)]%
        {beckmeyer2022recovering}
\bibfield{author}{\bibinfo{person}{Heiner Beckmeyer} {and} \bibinfo{person}{Timo Wiedemann}.} \bibinfo{year}{2022}\natexlab{}.
\newblock \showarticletitle{Recovering missing firm characteristics with attention-based machine learning}.
\newblock  (\bibinfo{year}{2022}).
\newblock


\bibitem[Bryzgalova et~al\mbox{.}(2025)]%
        {bryzgalova2025missing}
\bibfield{author}{\bibinfo{person}{Svetlana Bryzgalova}, \bibinfo{person}{Sven Lerner}, \bibinfo{person}{Martin Lettau}, {and} \bibinfo{person}{Markus Pelger}.} \bibinfo{year}{2025}\natexlab{}.
\newblock \showarticletitle{Missing financial data}.
\newblock \bibinfo{journal}{\emph{The Review of Financial Studies}} \bibinfo{volume}{38}, \bibinfo{number}{3} (\bibinfo{year}{2025}), \bibinfo{pages}{803--882}.
\newblock


\bibitem[Cahan et~al\mbox{.}(2023)]%
        {cahan2023factor}
\bibfield{author}{\bibinfo{person}{Ercument Cahan}, \bibinfo{person}{Jushan Bai}, {and} \bibinfo{person}{Serena Ng}.} \bibinfo{year}{2023}\natexlab{}.
\newblock \showarticletitle{Factor-based imputation of missing values and covariances in panel data of large dimensions}.
\newblock \bibinfo{journal}{\emph{Journal of Econometrics}} \bibinfo{volume}{233}, \bibinfo{number}{1} (\bibinfo{year}{2023}), \bibinfo{pages}{113--131}.
\newblock


\bibitem[Cao et~al\mbox{.}(2018)]%
        {cao2018brits}
\bibfield{author}{\bibinfo{person}{Wei Cao}, \bibinfo{person}{Dong Wang}, \bibinfo{person}{Jian Li}, \bibinfo{person}{Hao Zhou}, \bibinfo{person}{Lei Li}, {and} \bibinfo{person}{Yitan Li}.} \bibinfo{year}{2018}\natexlab{}.
\newblock \showarticletitle{Brits: Bidirectional recurrent imputation for time series}.
\newblock \bibinfo{journal}{\emph{Advances in neural information processing systems}}  \bibinfo{volume}{31} (\bibinfo{year}{2018}).
\newblock


\bibitem[Carhart(1997)]%
        {carhart1997persistence}
\bibfield{author}{\bibinfo{person}{Mark~M Carhart}.} \bibinfo{year}{1997}\natexlab{}.
\newblock \showarticletitle{On persistence in mutual fund performance}.
\newblock \bibinfo{journal}{\emph{The Journal of finance}} \bibinfo{volume}{52}, \bibinfo{number}{1} (\bibinfo{year}{1997}), \bibinfo{pages}{57--82}.
\newblock


\bibitem[Che et~al\mbox{.}(2018)]%
        {che2018recurrent}
\bibfield{author}{\bibinfo{person}{Zhengping Che}, \bibinfo{person}{Sanjay Purushotham}, \bibinfo{person}{Kyunghyun Cho}, \bibinfo{person}{David Sontag}, {and} \bibinfo{person}{Yan Liu}.} \bibinfo{year}{2018}\natexlab{}.
\newblock \showarticletitle{Recurrent neural networks for multivariate time series with missing values}.
\newblock \bibinfo{journal}{\emph{Scientific reports}} \bibinfo{volume}{8}, \bibinfo{number}{1} (\bibinfo{year}{2018}), \bibinfo{pages}{6085}.
\newblock


\bibitem[Chen and McCoy(2022)]%
        {chen2022missing}
\bibfield{author}{\bibinfo{person}{Andrew~Y Chen} {and} \bibinfo{person}{Jack McCoy}.} \bibinfo{year}{2022}\natexlab{}.
\newblock \showarticletitle{Missing values and the dimensionality of expected returns}.
\newblock \bibinfo{journal}{\emph{arXiv preprint arXiv:2207.13071}} (\bibinfo{year}{2022}).
\newblock


\bibitem[Chen and Chen(2022)]%
        {chen2022modeling}
\bibfield{author}{\bibinfo{person}{Elynn Chen} {and} \bibinfo{person}{Rong Chen}.} \bibinfo{year}{2022}\natexlab{}.
\newblock \showarticletitle{Modeling dynamic transport network with matrix factor models: with an application to international trade flow}.
\newblock \bibinfo{journal}{\emph{Journal of Data Science}} (\bibinfo{year}{2022}).
\newblock


\bibitem[Chen et~al\mbox{.}(2025)]%
        {chen2025distributed}
\bibfield{author}{\bibinfo{person}{Elynn Chen}, \bibinfo{person}{Xi Chen}, \bibinfo{person}{Wenbo Jing}, {and} \bibinfo{person}{Yichen Zhang}.} \bibinfo{year}{2025}\natexlab{}.
\newblock \showarticletitle{Distributed Tensor Principal Component Analysis with Data Heterogeneity}.
\newblock \bibinfo{journal}{\emph{Journal of the American Statistical Association, https://doi.org/10.1080/01621459.2025.2483481}} \bibinfo{number}{just-accepted} (\bibinfo{year}{2025}), \bibinfo{pages}{1--23}.
\newblock


\bibitem[Chen and Fan(2023)]%
        {chen2023statistical}
\bibfield{author}{\bibinfo{person}{Elynn Chen} {and} \bibinfo{person}{Jianqing Fan}.} \bibinfo{year}{2023}\natexlab{}.
\newblock \showarticletitle{Statistical inference for high-dimensional matrix-variate factor models}.
\newblock \bibinfo{journal}{\emph{J. Amer. Statist. Assoc.}} \bibinfo{volume}{118}, \bibinfo{number}{542} (\bibinfo{year}{2023}), \bibinfo{pages}{1038--1055}.
\newblock


\bibitem[Chen et~al\mbox{.}(2024a)]%
        {chen2024factor}
\bibfield{author}{\bibinfo{person}{Elynn Chen}, \bibinfo{person}{Jianqing Fan}, {and} \bibinfo{person}{Xiaonan Zhu}.} \bibinfo{year}{2024}\natexlab{a}.
\newblock \showarticletitle{Factor Augmented Matrix Regression}.
\newblock \bibinfo{journal}{\emph{arXiv preprint arXiv:2405.17744}} (\bibinfo{year}{2024}).
\newblock


\bibitem[Chen et~al\mbox{.}(2024b)]%
        {chen2024high}
\bibfield{author}{\bibinfo{person}{Elynn Chen}, \bibinfo{person}{Yuefeng Han}, {and} \bibinfo{person}{Jiayu Li}.} \bibinfo{year}{2024}\natexlab{b}.
\newblock \showarticletitle{High-Dimensional Tensor Classification with CP Low-Rank Discriminant Structure}.
\newblock \bibinfo{journal}{\emph{arXiv preprint arXiv:2409.14397}} (\bibinfo{year}{2024}).
\newblock


\bibitem[Chen et~al\mbox{.}(2024c)]%
        {chen2024highb}
\bibfield{author}{\bibinfo{person}{Elynn Chen}, \bibinfo{person}{Yuefeng Han}, {and} \bibinfo{person}{Jiayu Li}.} \bibinfo{year}{2024}\natexlab{c}.
\newblock \showarticletitle{High-Dimensional Tensor Discriminant Analysis with Incomplete Tensors}.
\newblock \bibinfo{journal}{\emph{arXiv preprint arXiv:2410.14783}} (\bibinfo{year}{2024}).
\newblock


\bibitem[Chen et~al\mbox{.}(2019)]%
        {chen2019constrained}
\bibfield{author}{\bibinfo{person}{Elynn Chen}, \bibinfo{person}{Ruey~S Tsay}, {and} \bibinfo{person}{Rong Chen}.} \bibinfo{year}{2019}\natexlab{}.
\newblock \showarticletitle{Constrained factor models for high-dimensional matrix-variate time series}.
\newblock \bibinfo{journal}{\emph{J. Amer. Statist. Assoc.}} (\bibinfo{year}{2019}).
\newblock


\bibitem[Chen et~al\mbox{.}(2024d)]%
        {chen2024semi}
\bibfield{author}{\bibinfo{person}{Elynn Chen}, \bibinfo{person}{Dong Xia}, \bibinfo{person}{Chencheng Cai}, {and} \bibinfo{person}{Jianqing Fan}.} \bibinfo{year}{2024}\natexlab{d}.
\newblock \showarticletitle{Semi-parametric tensor factor analysis by iteratively projected singular value decomposition}.
\newblock \bibinfo{journal}{\emph{Journal of the Royal Statistical Society Series B: Statistical Methodology}} (\bibinfo{year}{2024}), \bibinfo{pages}{qkae001}.
\newblock


\bibitem[Chen et~al\mbox{.}(2020)]%
        {chen2020modeling}
\bibfield{author}{\bibinfo{person}{Elynn Chen}, \bibinfo{person}{Xin Yun}, \bibinfo{person}{Rong Chen}, {and} \bibinfo{person}{Qiwei Yao}.} \bibinfo{year}{2020}\natexlab{}.
\newblock \showarticletitle{Modeling Multivariate Spatial-Temporal Data with Latent Low-Dimensional Dynamics}.
\newblock \bibinfo{journal}{\emph{arXiv preprint arXiv:2002.01305}} (\bibinfo{year}{2020}).
\newblock


\bibitem[CHEN et~al\mbox{.}(2020)]%
        {chen2020semiparametric}
\bibfield{author}{\bibinfo{person}{ELYNN~Y CHEN}, \bibinfo{person}{DONG XIA}, \bibinfo{person}{CHENCHENG CAI}, {and} \bibinfo{person}{JIANQING FAN}.} \bibinfo{year}{2020}\natexlab{}.
\newblock \showarticletitle{SEMIPARAMETRIC TENSOR FACTOR ANALYSIS BY ITERATIVELY PROJECTED SVD BY ELYNN Y. CHEN, DONG XIA, CHENCHENG CAI, AND JIANQING FAN}.
\newblock \bibinfo{journal}{\emph{arXiv preprint arXiv:2007.02404}} (\bibinfo{year}{2020}).
\newblock


\bibitem[Choi et~al\mbox{.}(2019)]%
        {choi2019s3}
\bibfield{author}{\bibinfo{person}{Dongjin Choi}, \bibinfo{person}{Jun-Gi Jang}, {and} \bibinfo{person}{U Kang}.} \bibinfo{year}{2019}\natexlab{}.
\newblock \showarticletitle{S3 CMTF: Fast, accurate, and scalable method for incomplete coupled matrix-tensor factorization}.
\newblock \bibinfo{journal}{\emph{PloS one}} \bibinfo{volume}{14}, \bibinfo{number}{6} (\bibinfo{year}{2019}), \bibinfo{pages}{e0217316}.
\newblock


\bibitem[De~Lathauwer et~al\mbox{.}(2000)]%
        {de2000multilinear}
\bibfield{author}{\bibinfo{person}{Lieven De~Lathauwer}, \bibinfo{person}{Bart De~Moor}, {and} \bibinfo{person}{Joos Vandewalle}.} \bibinfo{year}{2000}\natexlab{}.
\newblock \showarticletitle{A multilinear singular value decomposition}.
\newblock \bibinfo{journal}{\emph{SIAM journal on Matrix Analysis and Applications}} \bibinfo{volume}{21}, \bibinfo{number}{4} (\bibinfo{year}{2000}), \bibinfo{pages}{1253--1278}.
\newblock


\bibitem[Du et~al\mbox{.}(2023)]%
        {du2023saits}
\bibfield{author}{\bibinfo{person}{Wenjie Du}, \bibinfo{person}{David C{\^o}t{\'e}}, {and} \bibinfo{person}{Yan Liu}.} \bibinfo{year}{2023}\natexlab{}.
\newblock \showarticletitle{Saits: Self-attention-based imputation for time series}.
\newblock \bibinfo{journal}{\emph{Expert Systems with Applications}}  \bibinfo{volume}{219} (\bibinfo{year}{2023}), \bibinfo{pages}{119619}.
\newblock


\bibitem[Easton et~al\mbox{.}(2020)]%
        {easton2020attrition}
\bibfield{author}{\bibinfo{person}{Peter Easton}, \bibinfo{person}{Martin Kapons}, \bibinfo{person}{Peter Kelly}, {and} \bibinfo{person}{Andreas Neuhierl}.} \bibinfo{year}{2020}\natexlab{}.
\newblock \showarticletitle{Attrition bias and inferences regarding earnings properties; evidence from Compustat data}.
\newblock \bibinfo{journal}{\emph{Available at SSRN}} (\bibinfo{year}{2020}).
\newblock


\bibitem[Fama and French(1992)]%
        {fama1992cross}
\bibfield{author}{\bibinfo{person}{Eugene~F Fama} {and} \bibinfo{person}{Kenneth~R French}.} \bibinfo{year}{1992}\natexlab{}.
\newblock \showarticletitle{The cross-section of expected stock returns}.
\newblock \bibinfo{journal}{\emph{the Journal of Finance}} \bibinfo{volume}{47}, \bibinfo{number}{2} (\bibinfo{year}{1992}), \bibinfo{pages}{427--465}.
\newblock


\bibitem[Fama and French(1993)]%
        {fama1993common}
\bibfield{author}{\bibinfo{person}{Eugene~F Fama} {and} \bibinfo{person}{Kenneth~R French}.} \bibinfo{year}{1993}\natexlab{}.
\newblock \showarticletitle{Common risk factors in the returns on stocks and bonds}.
\newblock \bibinfo{journal}{\emph{Journal of financial economics}} \bibinfo{volume}{33}, \bibinfo{number}{1} (\bibinfo{year}{1993}), \bibinfo{pages}{3--56}.
\newblock


\bibitem[Freyberger et~al\mbox{.}(2025)]%
        {freyberger2025missing}
\bibfield{author}{\bibinfo{person}{Joachim Freyberger}, \bibinfo{person}{Bj{\"o}rn H{\"o}ppner}, \bibinfo{person}{Andreas Neuhierl}, {and} \bibinfo{person}{Michael Weber}.} \bibinfo{year}{2025}\natexlab{}.
\newblock \showarticletitle{Missing data in asset pricing panels}.
\newblock \bibinfo{journal}{\emph{The Review of Financial Studies}} \bibinfo{volume}{38}, \bibinfo{number}{3} (\bibinfo{year}{2025}), \bibinfo{pages}{760--802}.
\newblock


\bibitem[Freyberger et~al\mbox{.}(2020)]%
        {freyberger2020dissecting}
\bibfield{author}{\bibinfo{person}{Joachim Freyberger}, \bibinfo{person}{Andreas Neuhierl}, {and} \bibinfo{person}{Michael Weber}.} \bibinfo{year}{2020}\natexlab{}.
\newblock \showarticletitle{Dissecting characteristics nonparametrically}.
\newblock \bibinfo{journal}{\emph{The Review of Financial Studies}} \bibinfo{volume}{33}, \bibinfo{number}{5} (\bibinfo{year}{2020}), \bibinfo{pages}{2326--2377}.
\newblock


\bibitem[Grinold and Kahn(2000)]%
        {grinold2000active}
\bibfield{author}{\bibinfo{person}{Richard~C Grinold} {and} \bibinfo{person}{Ronald~N Kahn}.} \bibinfo{year}{2000}\natexlab{}.
\newblock \showarticletitle{Active portfolio management}.
\newblock  (\bibinfo{year}{2000}).
\newblock


\bibitem[Gu et~al\mbox{.}(2020)]%
        {gu2020empirical}
\bibfield{author}{\bibinfo{person}{Shihao Gu}, \bibinfo{person}{Bryan Kelly}, {and} \bibinfo{person}{Dacheng Xiu}.} \bibinfo{year}{2020}\natexlab{}.
\newblock \showarticletitle{Empirical asset pricing via machine learning}.
\newblock \bibinfo{journal}{\emph{The Review of Financial Studies}} \bibinfo{volume}{33}, \bibinfo{number}{5} (\bibinfo{year}{2020}), \bibinfo{pages}{2223--2273}.
\newblock


\bibitem[Han et~al\mbox{.}(2024)]%
        {han2024cp}
\bibfield{author}{\bibinfo{person}{Yuefeng Han}, \bibinfo{person}{Dan Yang}, \bibinfo{person}{Cun-Hui Zhang}, {and} \bibinfo{person}{Rong Chen}.} \bibinfo{year}{2024}\natexlab{}.
\newblock \showarticletitle{CP factor model for dynamic tensors}.
\newblock \bibinfo{journal}{\emph{Journal of the Royal Statistical Society Series B: Statistical Methodology}} \bibinfo{volume}{86}, \bibinfo{number}{5} (\bibinfo{year}{2024}), \bibinfo{pages}{1383--1413}.
\newblock


\bibitem[Jain and Oh(2014)]%
        {jain2014provable}
\bibfield{author}{\bibinfo{person}{Prateek Jain} {and} \bibinfo{person}{Sewoong Oh}.} \bibinfo{year}{2014}\natexlab{}.
\newblock \showarticletitle{Provable tensor factorization with missing data}.
\newblock \bibinfo{journal}{\emph{Advances in Neural Information Processing Systems}}  \bibinfo{volume}{27} (\bibinfo{year}{2014}).
\newblock


\bibitem[Jin et~al\mbox{.}(2021)]%
        {jin2021factor}
\bibfield{author}{\bibinfo{person}{Sainan Jin}, \bibinfo{person}{Ke Miao}, {and} \bibinfo{person}{Liangjun Su}.} \bibinfo{year}{2021}\natexlab{}.
\newblock \showarticletitle{On factor models with random missing: Em estimation, inference, and cross validation}.
\newblock \bibinfo{journal}{\emph{Journal of Econometrics}} \bibinfo{volume}{222}, \bibinfo{number}{1} (\bibinfo{year}{2021}), \bibinfo{pages}{745--777}.
\newblock


\bibitem[Kelly et~al\mbox{.}(2019)]%
        {kelly2019characteristics}
\bibfield{author}{\bibinfo{person}{Bryan~T Kelly}, \bibinfo{person}{Seth Pruitt}, {and} \bibinfo{person}{Yinan Su}.} \bibinfo{year}{2019}\natexlab{}.
\newblock \showarticletitle{Characteristics are covariances: A unified model of risk and return}.
\newblock \bibinfo{journal}{\emph{Journal of Financial Economics}} \bibinfo{volume}{134}, \bibinfo{number}{3} (\bibinfo{year}{2019}), \bibinfo{pages}{501--524}.
\newblock


\bibitem[Kolda and Bader(2009)]%
        {kolda2009tensor}
\bibfield{author}{\bibinfo{person}{Tamara~G Kolda} {and} \bibinfo{person}{Brett~W Bader}.} \bibinfo{year}{2009}\natexlab{}.
\newblock \showarticletitle{Tensor decompositions and applications}.
\newblock \bibinfo{journal}{\emph{SIAM review}} \bibinfo{volume}{51}, \bibinfo{number}{3} (\bibinfo{year}{2009}), \bibinfo{pages}{455--500}.
\newblock


\bibitem[Kong et~al\mbox{.}(2024)]%
        {kong2024teaformers}
\bibfield{author}{\bibinfo{person}{Linghang Kong}, \bibinfo{person}{Elynn Chen}, \bibinfo{person}{Yuzhou Chen}, {and} \bibinfo{person}{Yuefeng Han}.} \bibinfo{year}{2024}\natexlab{}.
\newblock \showarticletitle{TEAFormers: TEnsor-Augmented Transformers for Multi-Dimensional Time Series Forecasting}. In \bibinfo{booktitle}{\emph{The 34th International Joint Conference on Artificial Intelligence, 2025, AI for Time Series Workshop; arXiv preprint arXiv:2410.20439}}.
\newblock


\bibitem[Kozak et~al\mbox{.}(2020)]%
        {kozak2020shrinking}
\bibfield{author}{\bibinfo{person}{Serhiy Kozak}, \bibinfo{person}{Stefan Nagel}, {and} \bibinfo{person}{Shrihari Santosh}.} \bibinfo{year}{2020}\natexlab{}.
\newblock \showarticletitle{Shrinking the cross-section}.
\newblock \bibinfo{journal}{\emph{Journal of Financial Economics}} \bibinfo{volume}{135}, \bibinfo{number}{2} (\bibinfo{year}{2020}), \bibinfo{pages}{271--292}.
\newblock


\bibitem[Lettau(2024)]%
        {lettau20243d}
\bibfield{author}{\bibinfo{person}{Martin Lettau}.} \bibinfo{year}{2024}\natexlab{}.
\newblock \bibinfo{booktitle}{\emph{3D-PCA: Factor Models with Restrictions}}.
\newblock \bibinfo{type}{{T}echnical {R}eport}. \bibinfo{institution}{National Bureau of Economic Research}.
\newblock


\bibitem[Lewellen(2014)]%
        {lewellen2014cross}
\bibfield{author}{\bibinfo{person}{Jonathan Lewellen}.} \bibinfo{year}{2014}\natexlab{}.
\newblock \showarticletitle{The cross section of expected stock returns}.
\newblock \bibinfo{journal}{\emph{Forthcoming in Critical Finance Review, Tuck School of Business Working Paper}} \bibinfo{number}{2511246} (\bibinfo{year}{2014}).
\newblock


\bibitem[Light et~al\mbox{.}(2017)]%
        {light2017aggregation}
\bibfield{author}{\bibinfo{person}{Nathaniel Light}, \bibinfo{person}{Denys Maslov}, {and} \bibinfo{person}{Oleg Rytchkov}.} \bibinfo{year}{2017}\natexlab{}.
\newblock \showarticletitle{Aggregation of information about the cross section of stock returns: A latent variable approach}.
\newblock \bibinfo{journal}{\emph{The Review of Financial Studies}} \bibinfo{volume}{30}, \bibinfo{number}{4} (\bibinfo{year}{2017}), \bibinfo{pages}{1339--1381}.
\newblock


\bibitem[Liu and Moitra(2020)]%
        {liu2020tensor}
\bibfield{author}{\bibinfo{person}{Allen Liu} {and} \bibinfo{person}{Ankur Moitra}.} \bibinfo{year}{2020}\natexlab{}.
\newblock \showarticletitle{Tensor completion made practical}.
\newblock \bibinfo{journal}{\emph{Advances in Neural Information Processing Systems}}  \bibinfo{volume}{33} (\bibinfo{year}{2020}), \bibinfo{pages}{18905--18916}.
\newblock


\bibitem[Liu and Chen(2022)]%
        {liu2022identification}
\bibfield{author}{\bibinfo{person}{Xialu Liu} {and} \bibinfo{person}{Elynn Chen}.} \bibinfo{year}{2022}\natexlab{}.
\newblock \showarticletitle{Identification and estimation of threshold matrix-variate factor models}.
\newblock \bibinfo{journal}{\emph{Scandinavian Journal of Statistics}} (\bibinfo{year}{2022}).
\newblock


\bibitem[Mardani et~al\mbox{.}(2015)]%
        {mardani2015subspace}
\bibfield{author}{\bibinfo{person}{Morteza Mardani}, \bibinfo{person}{Gonzalo Mateos}, {and} \bibinfo{person}{Georgios~B Giannakis}.} \bibinfo{year}{2015}\natexlab{}.
\newblock \showarticletitle{Subspace learning and imputation for streaming big data matrices and tensors}.
\newblock \bibinfo{journal}{\emph{IEEE Transactions on Signal Processing}} \bibinfo{volume}{63}, \bibinfo{number}{10} (\bibinfo{year}{2015}), \bibinfo{pages}{2663--2677}.
\newblock


\bibitem[Montanari and Sun(2018)]%
        {montanari2018spectral}
\bibfield{author}{\bibinfo{person}{Andrea Montanari} {and} \bibinfo{person}{Nike Sun}.} \bibinfo{year}{2018}\natexlab{}.
\newblock \showarticletitle{Spectral algorithms for tensor completion}.
\newblock \bibinfo{journal}{\emph{Communications on Pure and Applied Mathematics}} \bibinfo{volume}{71}, \bibinfo{number}{11} (\bibinfo{year}{2018}), \bibinfo{pages}{2381--2425}.
\newblock


\bibitem[Mu et~al\mbox{.}(2014)]%
        {mu2014square}
\bibfield{author}{\bibinfo{person}{Cun Mu}, \bibinfo{person}{Bo Huang}, \bibinfo{person}{John Wright}, {and} \bibinfo{person}{Donald Goldfarb}.} \bibinfo{year}{2014}\natexlab{}.
\newblock \showarticletitle{Square deal: Lower bounds and improved relaxations for tensor recovery}. In \bibinfo{booktitle}{\emph{International conference on machine learning}}. PMLR, \bibinfo{pages}{73--81}.
\newblock


\bibitem[P{\'a}stor and Stambaugh(2003)]%
        {pastor2003liquidity}
\bibfield{author}{\bibinfo{person}{L'ubo{\v{s}} P{\'a}stor} {and} \bibinfo{person}{Robert~F Stambaugh}.} \bibinfo{year}{2003}\natexlab{}.
\newblock \showarticletitle{Liquidity risk and expected stock returns}.
\newblock \bibinfo{journal}{\emph{Journal of Political economy}} \bibinfo{volume}{111}, \bibinfo{number}{3} (\bibinfo{year}{2003}), \bibinfo{pages}{642--685}.
\newblock


\bibitem[Sun et~al\mbox{.}(2020)]%
        {sun2020low}
\bibfield{author}{\bibinfo{person}{Yiming Sun}, \bibinfo{person}{Yang Guo}, \bibinfo{person}{Charlene Luo}, \bibinfo{person}{Joel Tropp}, {and} \bibinfo{person}{Madeleine Udell}.} \bibinfo{year}{2020}\natexlab{}.
\newblock \showarticletitle{Low-rank Tucker approximation of a tensor from streaming data}.
\newblock \bibinfo{journal}{\emph{SIAM Journal on Mathematics of Data Science}} \bibinfo{volume}{2}, \bibinfo{number}{4} (\bibinfo{year}{2020}), \bibinfo{pages}{1123--1150}.
\newblock


\bibitem[Wen et~al\mbox{.}(2024)]%
        {wen2024tensor}
\bibfield{author}{\bibinfo{person}{Tao Wen}, \bibinfo{person}{Elynn Chen}, {and} \bibinfo{person}{Yuzhou Chen}.} \bibinfo{year}{2024}\natexlab{}.
\newblock \showarticletitle{Tensor-view Topological Graph Neural Network}. In \bibinfo{booktitle}{\emph{AISTATS, 2024, Val{\`e}ncia SPAIN}}.
\newblock


\bibitem[Wu et~al\mbox{.}(2025)]%
        {wu2025tensor}
\bibfield{author}{\bibinfo{person}{Yujia Wu}, \bibinfo{person}{Junyi Mo}, \bibinfo{person}{Elynn Chen}, {and} \bibinfo{person}{Yuzhou Chen}.} \bibinfo{year}{2025}\natexlab{}.
\newblock \showarticletitle{Tensor-fused multi-view graph contrastive learning}. In \bibinfo{booktitle}{\emph{Pacific-Asia Conference on Knowledge Discovery and Data Mining (PaKDD 2025)}}. Springer Nature Singapore, \bibinfo{pages}{16--28}.
\newblock


\bibitem[Wu et~al\mbox{.}(2024a)]%
        {wu2024conditional}
\bibfield{author}{\bibinfo{person}{Yujia Wu}, \bibinfo{person}{Bo Yang}, \bibinfo{person}{Elynn Chen}, \bibinfo{person}{Yuzhou Chen}, {and} \bibinfo{person}{Zheshi Zheng}.} \bibinfo{year}{2024}\natexlab{a}.
\newblock \showarticletitle{Conditional Prediction ROC Bands for Graph Classification}.
\newblock \bibinfo{journal}{\emph{AISTATS, 2025, Mai Khao, Thailand; https://proceedings.mlr.press/v258/wu25a.html}} (\bibinfo{year}{2024}).
\newblock


\bibitem[Wu et~al\mbox{.}(2024b)]%
        {wu2024conditionalb}
\bibfield{author}{\bibinfo{person}{Yujia Wu}, \bibinfo{person}{Bo Yang}, \bibinfo{person}{Yang Zhao}, \bibinfo{person}{Elynn Chen}, \bibinfo{person}{Yuzhou Chen}, {and} \bibinfo{person}{Zheshi Zheng}.} \bibinfo{year}{2024}\natexlab{b}.
\newblock \showarticletitle{Conditional Uncertainty Quantification for Tensorized Topological Neural Networks}.
\newblock \bibinfo{journal}{\emph{arXiv preprint arXiv:2410.15241}} (\bibinfo{year}{2024}).
\newblock


\bibitem[Zhou et~al\mbox{.}(2023)]%
        {zhou2023fast}
\bibfield{author}{\bibinfo{person}{Dan Zhou}, \bibinfo{person}{Ajim Uddin}, \bibinfo{person}{Zuofeng Shang}, \bibinfo{person}{Cheickna Sylla}, \bibinfo{person}{Xinyuan Tao}, {and} \bibinfo{person}{Dantong Yu}.} \bibinfo{year}{2023}\natexlab{}.
\newblock \showarticletitle{A Fast Non-Linear Coupled Tensor Completion Algorithm for Financial Data Integration and Imputation}. In \bibinfo{booktitle}{\emph{Proceedings of the Fourth ACM International Conference on AI in Finance}}. \bibinfo{pages}{409--417}.
\newblock


\end{thebibliography}

\appendix

\end{document}